\documentclass[aps,prl,twocolumn]{revtex4}
\usepackage{amsmath}
\usepackage{amsfonts}
\usepackage{amssymb}
\usepackage{graphicx}
\usepackage{bm}

\begin{document}
\title{On Biased Correlation Estimation}
\author{Thomas Sch\"urmann}
\thanks{Email: thomas.schuermann@dzbank.de}
\author{Ingo Hoffmann}
\affiliation{DZ BANK AG, Ludwig-Erhard-Allee 20, 40227 D\"usseldorf, Germany}
\begin{abstract}
In general, underestimation of risk is something which should be avoided as far as possible. Especially in financial asset management, equity risk is typically characterized by the measure of portfolio variance, or indirectly by quantities which are derived from it. Since there is a linear dependency of the variance and the empirical correlation between asset classes, one is compelled to control or to avoid the possibility of underestimating correlation coefficients. In the present approach, we formalize common practice and classify these approaches by computing their probability of underestimation. In addition, we introduce a new estimator which is characterized by having the advantage of a constant and controllable probability of underestimation. We prove that the new estimator is statistically consistent.
\end{abstract}
\date{\today}
\maketitle
\section{Introduction}
In statistics, the Pearson correlation coefficient, also referred to as the Pearson's $r$, is a measure of the linear correlation between two random variables $X$ and $Y$. It has a value between $+1$ and $-1$, where $1$ is total positive linear correlation, $0$ is no linear correlation, and $-1$ is total negative linear correlation. It was developed by Karl Pearson from a related idea introduced by Francis Galton in the 1880s. Twenty years after Galton conceived the idea, the correlation coefficient had found wide applications not only in biometry but also in experimental psychology and statistical economics. However, only rather few results on the properties of the finite sample estimator $r$ were known. Pearson and Filon \cite{PF} and Sheppard \cite{S} had proved that the large sample standard deviation of the estimator $r$ approaches $(1-\rho^2)/\sqrt{n}$, when $\rho$ is the true value.

In modern portfolio theory one is concerned with the estimation of variances of multi asset portfolios, which typically depend on the estimated correlation coefficients of normally distributed asset prices. In this situation it is appropriate to avoid underestimation of the total portfolio risk and one is compelled to take care that the correlation coefficients are suitably determined. Actually, one should be aware that the question of underestimation is independent whether the true correlation is negative or positive. Underestimation of correlations near +1 have the same relevance as correlations in the vicinity of -1.

To avoid underestimation, in practice \cite{Banks} a typical approach is to introduce a systematic bias by which the estimator $r$ is shifted towards +1. For instance, one widely used approach is to map the estimator $r$ to 0 if it is negative and leave it unchanged for $r>0$. Obviously, this kind of "brute force" transformation of $r$ leads to a new estimator, whose \textit{probability of underestimation} is significantly reduced for negative correlations. Another kind of typical transformation of $r$ is by adding a constant upward shift, and if the shifted value exceeds +1 then $r$ is put to +1.

As far as we know, the statistical properties of such concepts have never been explicitly discussed in literature so far. The reason might be that the standard approach is in mean-unbiased or median-unbiased estimators \cite{Br}\cite{SO}\cite{Fe}. This requirement seems to be accomplished for most purposes. However, in the present approach, the main interest is to introduce a systematic bias to take the practitioner's preference into account.

In the following section, we briefly describe the sampling probability density of $r$ which will be used extensively afterwards. Then, four different types of systematically biased correlation estimators are introduced. We consider their probability of underestimation, which is the main criterion for the classification in the present approach. A comparison of all estimators is given at the end.

\section{The probability density of correlation}

\begin{figure}[hbt]
\includegraphics[width=9.0cm,height=8.0cm]{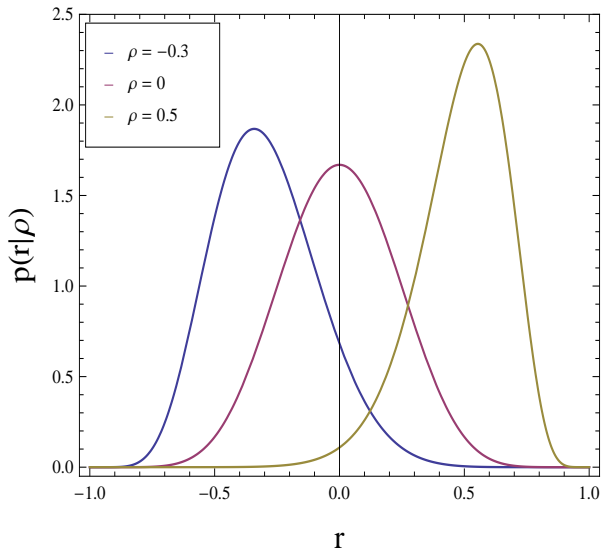}
\caption{Probability density $p(r|\rho)$ of the sample correlation estimator $r$, for samples of size $n=20$ and several values of $\rho$. The density is left/right-skewed for negative/positive values of $\rho$. For $\rho=0$, the density is symmetric (see text).} \label{fig1}
\end{figure}

In the following, let us consider independent and bivariate normally distributed random variables $(x_1,y_1),...,(x_n,y_n)$, with means $\mu_1,\mu_2$, variances $\sigma_1,\sigma_2$ and correlation coefficient $\rho$. The estimator $\hat r$ of $\rho$ is defined as
\begin{eqnarray}\label{r}
\hat r = \frac{\sum_{i=1}^n (x_i - \hat\mu_x)(y_i - \hat\mu_y)}{\sqrt{\sum_{i=1}^n (x_i - \hat\mu_x)^2\sum_{i=1}^n (y_i - \hat\mu_y)^2}},
\end{eqnarray}
while $\hat\mu$ is the mean of the data, i.e. $\hat\mu_x=1/n  \sum_{i=1}^n x_i$ and $\hat\mu_y=1/n  \sum_{i=1}^n y_i$.\\
\\
Now, the probability density function $p(r|\rho)$ of the estimator $\hat r$ is given by the following expression \cite{F}\cite{OP}
\begin{eqnarray}\label{p}
p(r|\rho) = \frac{2^{n-3}}{\pi\Gamma(n-2)}\,(1-\rho^2)^{(n-1)/2}(1-r^2)^{(n-4)/2}\nonumber\\
\times\sum_{k=0}^\infty\Gamma^2(\frac{n+k-1}{2})\,(2\,\rho\,r)^k,\qquad n>2,
\end{eqnarray}
while $\Gamma(x)$ is a special function called the Euler gamma function \cite{AS}. This distribution is valid for the case when the other parameters $\mu_1,\mu_2,\sigma_1$ and $\sigma_2$ are unknown but $\rho$ and $n$ are given (Fig.\,\ref{fig1}).
In the special case when $\rho=0$, the density can be simplified as
\begin{eqnarray}\label{p0}
p(r|0) = \frac{1}{\sqrt{\pi}}\,\frac{\Gamma(\frac{n-1}{2})}{\Gamma(\frac{n-2}{2})}\,(1-r^2)^{\frac{n}{2}-2}.
\end{eqnarray}

The expectation value of $r$ can be computed in terms of hypergeometric functions. However, the exact expression is rather cumbersome, but it can be shown \cite{SO} (Section 16.32) that
\begin{eqnarray}\label{Er}
\text{E}[r] = \rho \,\left(1-\frac{1-\rho^2}{2n}+\textit{O}\left(\frac{1}{n^2}\right)\right).
\end{eqnarray}
We can see that $\hat r$ is not a mean-unbiased estimator since $\text{E}[r]\neq \rho$, for finite values of $n$. But in the limit of large sample size $n\to\infty$, the expectation value tends to $\rho$, which implies that $\hat r$ is statistically consistent.

At this point, it should be mentioned that there exists a mean-unbiased estimator of the correlation coefficient which is a function of a complete sufficient statistic and is therefore the unique minimum variance unbiased estimator of $\rho$ \cite{OP}. However, here the criterion of mean-unbiasedness is not what we are looking for. Instead, our intention is to accept a systematic bias to get the freedom to reduce the \textit{probability of underestimation}. In literature, estimators whose probability of underestimation is equal to $1/2$ are called median-unbiased \cite{Br}. In the following section, we are looking for estimators whose probability of underestimation is even far less than $1/2$.

\section{Systematically biased estimators}
\begin{figure}[b]
\includegraphics[width=8.0cm,height=7.0cm]{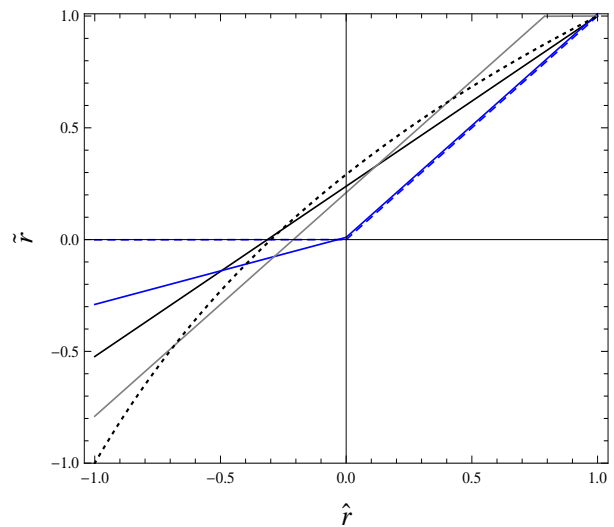}
\caption{The estimator $\tilde{r}_1$ for $a=0.23$ (black); $\tilde{r}_2$ for $b=0.7$ (blue); Of special interest is the degenerate case of $\tilde r_2$, for $b\to 1$. Then, $\tilde r_2$ is 0 for $r<0$ and unchanged otherwise (blue dashed). Parallel shift $\tilde r_3$ (gray). The new estimator $\tilde r$ is defined in (\ref{G}) (black dotted).} \label{fig2}
\end{figure}

For all $\hat r\in[-1,1]$ and constants $a,b,c\in[0,1)$, let us define four different types of biased estimators which are all functions of $\hat r$. The first three transformations are defined by (piecewise) linear transformations $G_i:\hat{r}\mapsto \tilde{r}_i$, $i=1,2,3$:
\begin{eqnarray}
\tilde r_1 &=& \,\,\,(1-a)\,\hat{r}+a, \label{r1}\\
\tilde r_2 &=&\begin{cases}(1-b)\,\hat r \qquad\qquad\hat r\in [-1,0)\\ \,\hat r \qquad\qquad\qquad\quad\hat r\in [\,0,1]\end{cases}\label{r2}\\
\tilde r_3 &=&\begin{cases}\hat r + c \,\,\,\qquad\quad\qquad\hat r\in [-1,1-c-\epsilon]\\ \frac{\epsilon\hat r+c}{\epsilon+c} \qquad\qquad\quad\,\,\,\,\hat r\in (1-c-\epsilon,1]\end{cases}\label{r3}
\end{eqnarray}
The constant $a$ in the first transformation is the intersection point of the vertical axis at $\hat r=0$, see Fig.\,\ref{fig2}. The map is linear on $[-1,1]$ and its image point at $\hat r=1$ is unchanged.

In the second transformation, the intersection point at the vertical axis at $\hat r=-1$ is given by $b-1$. Because of the kink at the origin, the map is only piecewise linear. Of special attention is the degenerate case for $b=1$. Here, $\hat r$ is mapped to 0, for all values $\hat r<0$, and is kept otherwise unchanged.

The third transformation is an upward-parallel shift with $c$ the intersection point of the vertical axis. This map is also defined piecewise while there is a kink at $\hat r=1-c-\epsilon$. As in the case before, this situation needs special attention.

As can be seen in Fig.\,\ref{fig2}, all of these transformations have the property that the support of their images do not entirely cover the set of all $\rho\in[-1,1]$. Therefore and for reasons discussed below, we introduce one more estimator $\tilde r$, which has the property to be a smooth and bijective transformation on $[-1,1]$. For its definition, let us briefly consider the notion of probability distributions by
\begin{eqnarray}\label{F}
\text{F}(x|\rho)= \int_{-1}^{x} dr\,p(r|\rho),
\end{eqnarray}
for given $x$ and $\rho$, and the density $p(r|\rho)$ is given by (\ref{p}). For any given $\rho$, it is the probability that the $r$ is smaller than $x$. The notion $\text{F}(x|\rho)$ is chosen to indicate its dependence on the parameter $\rho$. With this in mind, now we come to the following definition:\\
\\
\textbf{Definition.} For every fixed level of confidence $\alpha\in(0,1)$, consider the transformation $G_\alpha:[-1,1]\rightarrow[-1,1]$
\begin{eqnarray}\label{G}
\tilde{r}=G_\alpha(\hat r),
\end{eqnarray}
which is given by the solution of the integral equation
\begin{eqnarray}\label{int}
\int_{\hat r}^1\,dr\,p(r|\tilde r)=\alpha.
\end{eqnarray}
Applying (\ref{F}), definition (\ref{int}) can be rewritten by
\begin{eqnarray}\label{def}
\text{F}(\hat r|\tilde r)=1-\alpha.
\end{eqnarray}

The intention of (\ref{int}) is to determine the parameter $\tilde r$ such that the associated sampling with respect to the adjusted density $p(r|\tilde r)$ results with high probability in correlations which are above $\hat r$. The term "high probability" is quantified by the predefined value of $\alpha$. This approach, for instance, circumvents the disadvantages introduced by na\"{i}vely shifting the parameter of the distribution by a constant value such that the support of the density gets out of range $[-1,1]$.

\begin{figure}[t]
\includegraphics[width=8.0cm,height=7.0cm]{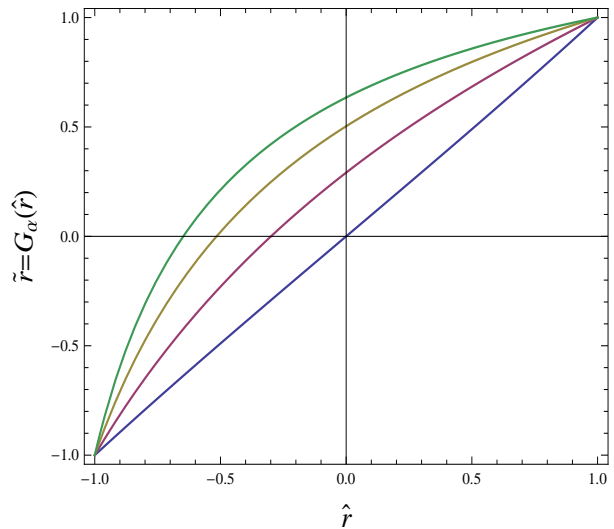}
\caption{Graph of the transformation $G_\alpha(r)$, for several certainties of overestimation $\alpha = 0.999, 0.99, 0.95, 0.5$ (from left to right) and samples of size 20. Values near $-1$ are typically stronger upward shifted than values near $+1$. For $\alpha=0.5$, we have $G_{1/2}(r)\approx r$.} \label{fig3}
\end{figure}

The probability distribution $\text{F}$ is continuous and strictly monotonic increasing. That implies that for every $\alpha\in(0,1)$, there is a unique estimator $\tilde r$ which is a smooth function with respect to $\hat r$. In addition, the boundary points $\hat r=\pm 1$ are mapped to the image $\tilde r=\pm 1$, for all $\alpha\in(0,1)$. A typical case ($\alpha = 0.95$) is shown in Fig.\,\ref{fig2} (black dotted) and for several levels $\alpha$ in Fig.\,\ref{fig3}. The upward shifts of $\hat{r}$ near $-1$ are stronger than for estimates near $+1$. For the specific case $\alpha\to 1$, $\tilde r$ jumps from $-1$ to $+1$ at $\hat r=-1$, such that in this limit we have $\tilde r=1$ for all $\hat r\in(-1,+1]$.

For an entire classification of (\ref{r1})-(\ref{r3}) and (\ref{G}), we next introduce the precise notion of how to measure the probability of underestimation.
\section{The Probability of Underestimation}
\begin{figure}[t]
\includegraphics[width=8.0cm,height=7.0cm]{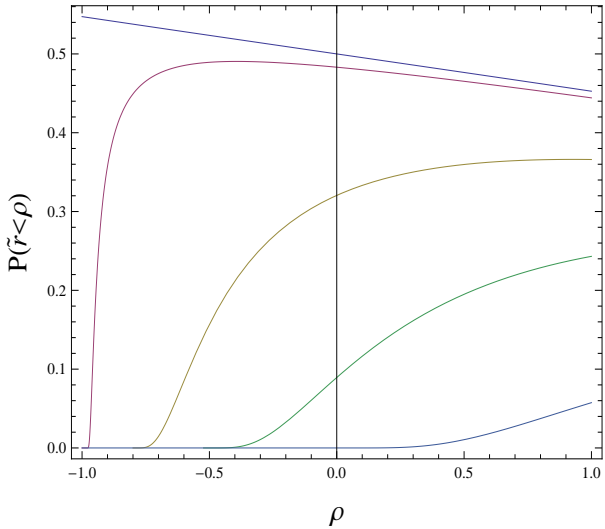}
\caption{Probability of underestimation for the estimator (\ref{r1}) with $a=0.0, 0.01, 0.05, 0.1, 0.238, 0.5$ (from top to bottom) and sample size $n=20$. The case $a=0$ corresponds to $\tilde r_1\equiv\hat r$. For increasing $a$, the probability of underestimation is decreasing.} \label{fig4}
\end{figure}

\begin{figure}[t]
\includegraphics[width=8.0cm,height=7.0cm]{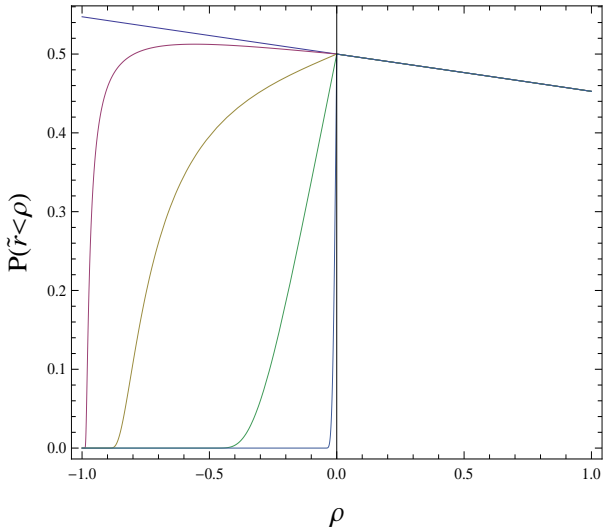}
\caption{Probability of underestimation for (\ref{r2}) with $b=0.0, 0.01, 0.05, 0.1, 0.5, 0.95$ (from top to bottom) and sample size $n=20$. The case $b=0$ corresponds to $\tilde r_2\equiv\hat r$. For $b\to 1$ the shape of probability at the origin tends to be a step function.} \label{fig5}
\end{figure}

The probability of underestimation is a quantity to specify the chance by which an estimate of a sample is less than the parameter under consideration. For its definition, the parametric sample distribution of the estimator has to be known. With the notion of (\ref{F}), in our context the probability of underestimation is formally given by
\begin{eqnarray}\label{Pr}
\text{P}(\tilde r<\rho)=\text{F}(G^{-1}(\rho)|\rho).
\end{eqnarray}
The right-hand side is obtained by applying the measure transformation corresponding to the maps (\ref{r1})-(\ref{r3}) or (\ref{G}) respectively. For its explicit computation, we have to determine the corresponding inverse map $G^{-1}$. Therefore, let us start with \\
\\
\underline{Case 1:} For the linear map in (\ref{r1}), the inverse map is given by
\begin{eqnarray}\label{i1}
G^{-1}_1(\tilde r_1)&=&\frac{\tilde r_1-a}{1-a},\,\qquad\tilde r_1\in [2a-1,1],
\end{eqnarray}
with $a\in[0,1)$. In Fig.\ref{fig4}, we see the probability $(\ref{Pr})$ for several values of $a$. For $a=0$, the ordinary case of $\tilde r_1=\hat r$ is reproduced. For increasing $a$, the probability of underestimation decreases. Of course it is equal to zero for $\tilde r<2a-1$.\\
\\
\underline{Case 2:} For the estimator (\ref{r2}), the inverse map has to be considered piecewise and is given by
\begin{eqnarray}\label{i2}
G^{-1}_2(\tilde r_2)&=&\begin{cases}\frac{\tilde r_2}{1-b}\,\,\,\qquad\qquad\tilde r_2\in [b-1,0)\\
\,\tilde r_2\quad\qquad\qquad\,\tilde r_2\in [0 ,1] \end{cases}
\end{eqnarray}
for $b\in[0,1)$. Of special interest is the degeneracy for $b\to 1$. Then, the probability weight of the domain $\tilde r<0$ shrinks to the origin with an increasing peak near zero such that the probability weight at zero stays finite. The corresponding probability of underestimation is shown in Fig.\,\ref{fig5}.
For the case $b\to 1$, the probability distribution becomes discontinuous and approaches a step-function at $\rho=0$. In Fig.\,\ref{fig5}, this case is illustrated for $b=0.95$. In the domain of $\rho\geq 0$, the probability of underestimation is the same for all $b\in[0,1)$ because $\tilde r_2$ is identical to $\hat r$.  \\
\begin{figure}[t]
\includegraphics[width=8.0cm,height=7.0cm]{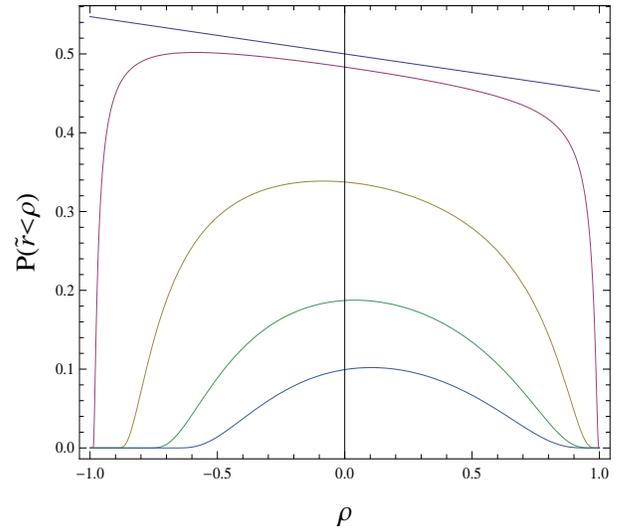}
\caption{Probability of underestimation for (\ref{r3}) with $c=0.0, 0.01, 0.1, 0.21, 0.3$ (from top to bottom) and sample size $n=20$. The case $c=0$ corresponds to $\tilde r_3\equiv\hat r$. All lines are for $\epsilon\to 0$.} \label{fig6}
\end{figure}
\\
\setlength{\parindent}{0pt}\underline{Case 3:} The inverse of the transformation (\ref{r3}) is given by
\begin{eqnarray}\label{i3}
G^{-1}_3(\tilde r_3)=\begin{cases}\,\tilde r_3-c\qquad\qquad\qquad\,\tilde r_3\in [c-1,1-\epsilon]\\
(1+\frac{c}{\epsilon})\,\tilde r_3-\frac{c}{\epsilon}\,\,\,\qquad\,\,\tilde r_3\in (1-\epsilon,1].
\end{cases}
\end{eqnarray}
For the computation of the probability of underestimation, we have to take into account that the limit $\epsilon\to 0$ has to be performed after integration has been computed. The result can be seen in Fig.\,\ref{fig6}. For increasing upward shifts the probability of underestimation becomes increasingly smaller. \\
\setlength{\parindent}{12pt}

All of the previous three cases have in common that their probability of underestimation is not constant for $\rho\in[-1,1]$, but is a strongly varying quantity. On the other hand, for the estimator defined in (\ref{G}), we have the following statement: \\
\\
\textbf{Theorem 1.} Given $\alpha\in(0,1)$. Let $\tilde r=G_\alpha(\hat r)$ be the estimator defined in (\ref{G}). Then, for every sample size $n>2$, the probability of underestimation (\ref{Pr}) is constant and given by
\begin{eqnarray}\label{P}
\text{P}(\tilde{r}<\rho)=1-\alpha.
\end{eqnarray}
\textbf{Proof.} By assumption, the map $G_\alpha$ is continuous and strictly monotonic increasing. Therefore, $G_\alpha$ is invertible. Let the inverse of $G_\alpha$ be $G_\alpha^{-1}$. Since $G_\alpha$ is smooth, the derivative of $G_\alpha$ with respect to $\hat r$ exists on $[-1,1]$. By applying the measure transformation corresponding to $d\tilde{r}/d\hat{r}=G^{'}_\alpha(\hat r)$, we can rewrite the left-hand side of (\ref{P}) by
\begin{eqnarray}\label{proof}
\text{P}(\tilde{r}<\rho)=\text{F}(G_\alpha^{-1}(\rho)|\rho).
\end{eqnarray}
By definition (\ref{def}), the right-hand side of (\ref{proof}) is equal to $1-\alpha$.$\hfill\Box$\\
\\
The theorem obviously shows the advantage of (\ref{G}) compared to the other estimators discussed above. In addition, we have\\
\\
\textbf{Theorem 2.} The estimator $\tilde r$ defined in (\ref{def}) is statistically consistent.\\
\\
\textbf{Proof.} To prove consistency of $\tilde r$ with respect to (\ref{def}), we consider that the ordinary Pearson $\hat r$ is already known to be a statistically consistent estimator. Therefore, it is sufficient to show that $\tilde r\to\hat r$, for $n\to\infty$. For large $n$, the probability density (\ref{p}) approaches asymptotically to a normal distribution \cite{SO}\cite{Fe}, with mean $\rho$ and standard deviation $(1-\rho^2)/\sqrt n$. In this case, the integral on the left-hand side in (\ref{int}) can be performed. Since for every given real number $\epsilon>0$, there exists a positive integer $N$, such that for all $n>N$, we have
\begin{eqnarray}\label{lim}
\Big|\,\text{F}(\hat r|\tilde r)-\frac{1}{2}\,\text{erfc}\big(\sqrt{\frac{n}{2}}\,\frac{\tilde r-\hat r}{1-\tilde r^2}\big)\Big|<\epsilon.
\end{eqnarray}
The function $\text{erfc}(x)$ is related to the ordinary Gaussian error integral \cite{AS} by $\text{erfc}(x)=1-\text{erf}(x)$. Now, because of the definition (\ref{def}), we can replace the distribution function in (\ref{lim}) by $1-\alpha$. Then, the left-hand side of (\ref{lim}) can be equated to zero and after some simple algebraic manipulations we obtain the following condition in terms of the Gaussian error function:
\begin{eqnarray}\label{erf}
\text{erf}\big(\sqrt{\frac{n}{2}}\,\frac{\tilde r-\hat r}{1-\tilde r^2}\big)=2\,\alpha-1.
\end{eqnarray}
For every $\alpha\in(0,1)$, let $q_\alpha$ be the real valued solution of the equation $\text{erf}(q_\alpha/\sqrt 2)=2\,\alpha-1$. Then, (\ref{erf}) can be written equivalently as
\begin{eqnarray}\label{sol}
\hat r=\tilde r - \,q_\alpha\,\frac{1-\tilde r^2}{\sqrt n}.
\end{eqnarray}
For the asymptotic case $n\to \infty$, this expression becomes exact and we find that $\tilde r \to\hat r$. This proves the consistency of the estimator $\tilde r$.$\hfill\Box$
\\
\\
It should be mentioned here that although the relation (\ref{sol}) is only an approximation when $n$ is finite, this equation works pretty well already, even for small size samples of about $n\geq 20$. For practical purposes, we therefore explicitly write down the large sample approximation for the map $\tilde r=G_\alpha(\hat r)$ in (\ref{G}), that is
\begin{eqnarray}\label{Ga}
\tilde r=\frac{\sqrt{n+4 q_\alpha\sqrt{n}\,\hat r+4\,q_\alpha^2  }-\sqrt{n}}{2 q_\alpha},
\end{eqnarray}
for all $n>4q_\alpha^2$. This also confirms the limit behavior for $n\to\infty$.

\section{Summary}
Common approaches of biased correlation estimation in financial risk management have been formalized by piecewise linear transformations. Based on this formalization, we discussed their corresponding probability of underestimation. As a result, we found that this probability is a strongly varying quantity depending on the value of the parameter under consideration. To resolve that problem, a new correlation estimator has been introduced. The most important property of this estimator is that its corresponding probability of underestimation is constant, and in addition, a controllable quantity. For the new estimator, we also proved statistical consistency.
\acknowledgments

\newpage{}

\begin{thebibliography}{99}
\bibitem{Banks} For reasons of data privacy, the names of the financial services providers or financial institutions using these approaches will not be mentioned here.
\end{thebibliography}
\end{document}